\documentclass[draft]{birkjour}
\usepackage[noadjust]{cite}
\usepackage{xcolor}
\RequirePackage[all]{xy}

\usepackage{bbold}
\usepackage{amsmath}
\usepackage{amsthm}
\usepackage{amssymb}
\usepackage{amsfonts}
\usepackage{tensor}
\usepackage{braket}
\usepackage{slashed}

\theoremstyle{definition}

\theoremstyle{remark}

\numberwithin{equation}{section}

\newcommand{\BibTeX}{B\kern-0.1emi\kern-0.017emb\kern-0.15em\TeX}
\newcommand{\XYpic}{$\mathrm{X\kern-0.3em\raisebox{-0.18em}{Y}}$-$\mathrm{pic}\,$}

\newcommand{\cl}{C \kern -0.1em \ell}  

\newcommand{\vect}{\textbf}

\newtheorem{theorem}{Theorem}[section]

\newcommand{\f}{\vect{f}}
\newcommand{\g}{\vect{g}}
\newcommand{\x}{\vect{x}}

\newcommand{\A}{\vect{A}}
\newcommand{\vd}{\boldsymbol{\partial}}
\newcommand{\vn}{\boldsymbol{\nabla}}
\newcommand{\tht}{\boldsymbol{\theta}}

\newcommand{\ed}{\end{document}}
\begin{document}

%
%
%
%
%
%
%
%
%

\title[Self-Dual Maxwell Fields from Clifford Analysis]
 {Self-Dual Maxwell Fields from Clifford Analysis}
\author[C.J.Robson]{C.J.Robson}

\address{%
London School of Economics\\
UK}
\email{c.j.robson@lse.ac.uk}

\date{\today}
\dedicatory{Published in Adv. Appl. Clifford Algebras (2025) 35:7}
\begin{abstract}
	The study of complex functions is based around the study of holomorphic functions, satisfying the Cauchy-Riemann equations. The relatively recent field of Clifford Analysis  lets us extend many results from Complex Analysis to higher dimensions.  In this paper,  I decompose the Cauchy-Riemann equations for a general Clifford algebra into grades using the Geometric Algebra formalism,  and show that for the Spacetime Algebra $Cl(3,1)$ these equations are the equations for a self-dual source free Maxwell field, and for a massless uncharged Spinor.  This shows a deep link between fundamental physics and the Clifford geometry of Spacetime. 
\end{abstract}

\label{page:firstblob}
\maketitle

\section{Introduction and Background}
This paper is about the application of Clifford (or Geometric) algebras to physics. In particular, it combines two recent approaches to Clifford algebras. The first is Geometric Algebra \cite{Hestenes1987}\cite{Gull1993}\cite{Doran2003}, which is an approach to Clifford algebras using a geometrical representation described below. The second is Clifford Analysis \cite{Sommen1984}\cite{Ryan2003}, which extends the results of Complex Analysis first to the Quaternions, and then to a general Clifford algebra. I shall discuss both of these below. I am aware of one other work combining these approaches, the PhD Thesis \cite{Roberts2022}. \\
In general, the Clifford approach to geometry is based upon the ideas of W.K.Clifford and H.G. Grassmann \cite{Lounesto2009}, and involves giving a direct geometric representation of the n-dimensional subspaces of a Clifford algebra in terms of the n-dimensional linear subspaces of a geometric space, for example lines, planes, volumes, and so on. The Clifford algebra on a surface is also isomorphic as vector spaces to the Differential Forms on the same surface. This is what allows us to use techniques from the cohomology theory of forms in the Clifford Algebra context \cite{Roberts2022}. In this paper I adopt the viewpoint that, just as the complex plane $\mathbb{C}^{2}$ can be seen as the real plane $\mathbb{R}^{2}$ equipped with a multiplication, and hence promoted from a vector space to an algebra. The Clifford Algebra $Cl(pq,)$ is likewise the algebra generated from the vector space $\mathbb{R}^{p,q}$ equipped with a suitable product. \\ 
A more technical distinction  between Clifford geometry and the standard approach  is given by Dieudonne as follows \cite{Dieudonne1979}. In standard approaches,we consider both a vector space $\mathcal{V}$, and its dual $\mathcal{V}^{\star}$. We cannot identify these two spaces, but we can associate elements in $\mathcal{V}$ to elements in $\mathcal{V}^{\star}$ if we define an inner product between these spaces. Then a $p$-dimensional subspace $V_{p}\in\mathcal{V}$ will be dual to an $(n-p)$-dimensional subspace $W_{n-p}\in\mathcal{V}^{\star}$ iff $(v,w)=0$ for all $v\in V_{p}$ and $w\in W_{n-p}$, where $n$ is the total dimension of $\mathcal{V}$ and $\mathcal{V}^{\star}$, and $(\ , \ )$ denotes the inner product. \\
Conversely, in the Clifford approach to Geometry, we only have one space, $\mathcal{V}$, which is equipped with a product to turn it into an algebra; in exact analogy to the transition between $\mathbb{R}^{2}$ and $\mathbb{C}$.  More details are given below, but the essential point is that the dual of a $p$-dimensional subspace $V_{p}$ is given by its $(n-p)$ dimensional complement in $\mathcal{V}$ itself. This has advantages for physics; first of all by cutting down on the mathematical entities required, and secondly because most objects in physics can be given a direct geometric interpretation in this construction (See \cite{Doran2003} for examples). \\
\subsection{Geometric Algebras}
I begin the paper with some definitions and notation. First, I will review the Geometric Algebra formalism for Clifford algebras. Then, I will discuss Clifford Analysis as a generalisation of Complex analysis.  This section is mainly based on \cite{Macdonald2010} and \cite{Doran2003}. Other excellent references are \cite{Chisholm2012}\ and \cite{Dorst2009}.\\
An (orthogonal) Clifford Algebra $Cl(p,q)$ with a non-degenerate metric is described by $n=p+q$ generators $e_{i}$, satisfying
\begin{equation}
	e_{i}e_{j}-e_{j}e_{i}=\eta_{ij}
\end{equation}
Where $\eta_{ii}$ is 1 for $i$ from 1 to $p+1$, $\eta_{ii}=-1$ for $i$ from $p+1$ to $n$, and $\eta_{ij}=0$ otherwise.\\ 
Clifford algebras are a graded algebra, with the grading given by the number of generators that are multipled togther. For example, $e_{i}$ is grade 1, $e_{i}e_{j}e_{k}$ is grade 3, and so on. \\
Returning to the general case, we can extract the grade $i$ elements of a Clifford algebra element $X $, and we denote this $\langle X\rangle_{i}$. An element which contains only objects of the same grade is called a blade. \\
A Clifford algebra can be split into even and odd parts, corresponding to collecting the blades with even and odd grade. The even part is called the Spin algebra, denoted $Spin(p,q)$. \cite{Lawson1990}.  \\
In the Geometric Algebra (GA) representation of Clifford Algebras, we take each grade to describe an $k$- dimensional subspace of $\mathbb{R}^{n}$. The highest grade object is denoted $I$, and describes a $n$-volume. So in 3 dimensions, the scalar describes a point, $e_{i}$ describes a line, $e_{i}e_{j}$ describes a plane, and the pseudoscalar $I=e_{1}e_{2}e_{3}$ describes a 3-volume.\\
In this paper, we will be especially interested in the Clifford Algebra $Cl(3,1)$, which is generated by Minkowski Spacetime.  It consists of 1 scalar, 4 vectors $e_{a}$, six bivectors $e_{a}e_{b}$, 4 trivectors $e_{a}e_{b}e_{c}$, and 1 pseudoscalar $I=e_{0}e_{1}e_{2}e_{3}$.\\
We use the index $0$ to refer to the timelike direction, for which $e_{0}^{2}=-1$, and  $i,j,k = \{1,2,3\}$ for spacelike directions, for which $e_{i}^{2}=1$.  
These correspond to the gamma matrices $\gamma_{a}$ in the usual Dirac theory, and the pseudoscalar $I$ corresponds to $\gamma_{5}\equiv\gamma_{0}\gamma_{1}\gamma_{2}\gamma_{3}$. \cite{Gull1993} \\
It is also worth noting that the bivector part of the algebra corresponds to Lorentz rotations \cite{Doran2003}. The timelike rotations correspond to terms of the form $e_{0}e_{i}$, and spacelike rotations correspond to $e_{i}e_{j}$. \\
When we multiply two blades  $\langle A \rangle_{i}$ and $\langle B \rangle_{j}$, the lowest grade object we can form has grade $\lvert i-j\rvert$, and is called the dot product, denoted by $A\cdot B$ or $\langle A\cdot B\rangle_{\lvert i-j\rvert}$. The highest grade element has grade $i+j$, and is called the wedge product, denoted $A\wedge B$ or $\langle A\wedge B\rangle_{i+j}$. Note that these may be identically zero (for example if $i+j>n$), and that in general one or both will be. For vectors $a,b$ these definitions become
\begin{align}\nonumber
	a\cdot b=ab-ba\\
	a\wedge b=ab-ba
\end{align}  \\
Duality relations exist between blades of grades $k$ and $n-k$. These are defined in various ways in the literature, which differ up to a sign. I shall use the convention
\begin{equation}
	A^{\star}=AI^{-1}
\end{equation}
where $A^{\star}$ is the dual of $A$. If $A$ has grade $k$, then $A^{\star}$ has grade $n-k$. The following property of the dual will be important later
\begin{align}\nonumber
	A\wedge BI^{-1}=\big(A\wedge B\big)I^{-1}\\
	A\cdot BI^{-1}=\big(A\cdot B\big)I^{-1}
\end{align}
In comparison to the usual vector calculus,  $a\cdot b$ is the usual dot product, and in 3d we have
\begin{equation}
	a\wedge b \ = \ (a \times b)I^{-1}
\end{equation}
where $a \times b$ is the cross product. This formula just means that (in 3 dimensions) $a\wedge b$ is the area between $a$ and $b$, and $a \times b$ is the length orthogonal to that area. Note, however,  that we can define $a\wedge b$ in any dimension, not just in 3d.  \\
We call an element made up of different grades a Multivector. Then the most general function we can write from $Cl(p,q)\rightarrow Cl(p,q)$ is
\begin{equation}
	\vect{z}=f_{0}+f_{i}e_{i}+f_{ij}e_{i}e_{j}+...+f_{1...d}e_{1}...e_{d}
\end{equation}
where the $f$ are scalar functions. In this paper I shall only consider functions of $\x$, where $\x=x_{i}e_{i}$ is the usual position vector. \\
Next, we define a vector derivative
\begin{equation}\label{eqn:vecder}
	\vd=\frac{\partial}{\partial\x}\equiv \ e_{i}\frac{\partial}{\partial x}
\end{equation}
Once we have this, we can define the wave/laplacian operator as $\vd^{2}$. Applying the vector derivative to a blade will either raise or lower the grades of the components of multivector by 1. In general it will do both. We therefore write
\begin{equation}
	\vd\vect{z}=\vd\cdot\vect{z}+\vd\wedge\vect{z}
\end{equation}
where   $\vd\cdot\vect{z}$ and $\vd\wedge\vect{z}$ are the grade lowering and grade raising parts respectively. \\
\subsection{Clifford Analysis}
The introduction of the vector derivative brings us back to Clifford Analysis. Clifford Analysis is focused around finding and understanding solutions of the Dirac eqation $\vd\phi=0$.. This means that Clifford Analysis intersects with other areas of mathematics, for example Spin Geometry, in particular index theory \cite{Lawson1990} \\
To show the link between Clifford Algebras and Complex Analysis, consider the algebra $Cl(2)$. This consists of a scalar, two vectors $e_{1}$ and $e_{2}$, and a bivector pseudoscalar, $I=e_{1}e_{2}$. The even subalgebra, $Spin(2)$, is spanned by $\{1, I\}$, with $I^{2}=-1$. Therefore $Spin(2)\cong\mathbb{C}$. A generic $Spin(2)$-valued function is given by $u(z)+Iv(z)$, where $z=x+Iy$. Equivalently, we can consider the variable $\tilde{z}=e_{1}z=xe_{1}+ye_{2}$. Then
\begin{align}\nonumber
	\vd\big(u(\tilde{z}))+Iv(\tilde{z})\big)=e_{1}\frac{\partial u}{\partial x}+e_{2}\frac{\partial u}{\partial y}+e_{1}I\frac{\partial v}{\partial x}+e_{2}I\frac{\partial v}{\partial y}\\
	=e_{1}\Big(\frac{\partial u}{\partial x}-\frac{\partial v}{\partial y}\Big)+e_{2}\Big(\frac{\partial u}{\partial y}+\frac{\partial{v}}{\partial x}\Big)
\end{align}
Therefore the condition $\vd(u+Iv)=0$ is equivalent to the pair of equations
\begin{equation}
	u_{x}=v_{y}; \ \ u_{y}=v_{x}
\end{equation}
We can recognise these as the Cauchy- Riemann equations, and a complex function satisfying $\vd f=0$ is known as a holomorphic function. In Clifford Analysis, solutions to the equation $\vd\f=0$ are usually called Monogenic functions, though they are occasionally known as Clifford Holomorphic functions. or Regular functions in older literature. I shall refer to them as Monogenic functions in this paper. \\ 
An important property of the Cauchy-Riemann equations is that any solution to them is also a harmonic function, i.e. $\vd^{2}\vect{z}=0$.  This also applies to Clifford analysis, where we have  \cite{Delanghe1992} \\
\begin{theorem}\label{thm:main}
	$Mon_{p}(\mathcal{M})$,the vector space of monogenic functions of degree {p} over a space $\mathcal{M}$, is identical to $Har_{p}(\mathcal{M})$, the vector space of harmonic forms of degree $p$. 
\end{theorem}
Clifford Analysis has its beginnings in the work of Moisil, Théodoresco and Riesz  \cite{Moisil1931} \cite{Riesz1958} in the 1930s, which successfully extended the key results of Complex Analysis to the quaternionic case. R. Fueter then extended this to higher dimensions via Clifford Algebras\cite{Fueter1935}\cite{Fueter1948}. In the late 20th century, this was developed by figures like Hestenes, Sobcyck and Delenghe \cite{Hestenes1987}\cite{Delanghe1992}with results like the Cauchy Integral theorem and the Cauchy- Kowalseka extension being extended from the complex numbers to general Clifford Algebras. For a good overview, see the paper \cite{Ryan2003}.  \\
. \subsection{Maxwell's Equations and Clifford Algebras}
This paper will ultimately show a link between Maxwell's equations and the Dirac equation for multivector functions defined on the Clifford Algebra $Cl(3,1)$. I will therefore briefly review the formulation of Maxwell's equations in the Clifford Algebra context, summarising the account in \cite{Doran2003}. \\
In natural units, the source-- free Maxwell equations in Minkowski space can be written as
\begin{align}\nonumber
	\vn\cdot\vect{E}&=0\\ \nonumber
	\vn\times\vect{E}&=-\frac{\partial}{\partial t}\vect{B} \\ \nonumber
	\vn\cdot\vect{B}&=0 \\ 
	\vn\times\vect{B}&=\frac{\partial}{\partial t}\vect{E}
\end{align}
where $\times$ is the usual cross product in $\mathbb{R}^{3}$ and the $\vect{E}$ and the $\vect{B}$ are three dimensional vectors (in the spacelike directions$ \{1,2,3\}$) and $\vn$ is the 3d vector derivative in the spacelike directions. When we go from Minkowski space to the Clifford Algebra $Cl(3,1)$, these equations become
\begin{align}\nonumber
	\vn\cdot\vect{E}&=0\\ \nonumber
	\vn\wedge\vect{E}&=-\frac{\partial}{\partial t}(I\vect{B}) \\ \nonumber
	\vn\cdot\vect{B}&=0 \\ 
	\vn\times\vect{B}&=\frac{\partial}{\partial t}(I\vect{E})
\end{align}
These are mathematically identical to the standard vector calculus equations, merely expressed in a different formalism. We can take advantage of the freedom to add elements of different grades to write
\begin{equation}	
	\vn\vec{E}=\vn\cdot\vect{E}+\vn\wedge\vect{E}=-\partial_{t}\big(I\vect{B}\big)
\end{equation}
and similarly 
\begin{equation}
	\vn\big(I\vect{B}\big)=-\partial_{t}\vect{E}
\end{equation}
We combine these  to get
\begin{equation}\label{eqn:Maxwell2}
	\vn(\vect{E}+I\vect{B})+\partial_{t}(\vect{E}+I\vect{B})=0
\end{equation}
Here $\vect{E}$ is a vector, and $I\vect{B}$ is a dual trivector. It is more natural in the Clifford Algebra setting to write them both as bivectors. To do this, we multiply them by $\gamma_{0}$ (and drop the boldface type) so $E=E_{i}\gamma_{0}\gamma_{i}$ and $B=B_{i}\gamma_{0}\gamma_{i}$. Therefore $I\vect{B}=B_{i}\epsilon_{ijk}\gamma_{j}\gamma_{k}$\footnote{There is nothing stopping us with working directly with $\tilde{B}\equiv IB $ directly. It is a matter of notational preference. Here I am following the derivation in \cite{Doran2003} and so I shall use their conventions}.
Now we can define the electromagnetic field strength as
\begin{equation}
	F=E+IB
\end{equation}
Then equation (\ref{eqn:Maxwell2}) becomes
\begin{equation}
\vd F=0
\end{equation}
As in the vector calculus case, the fact that $\vd\wedge F=0$ means we can write 
\begin{equation}
	F=\vd\wedge\vect{A}
\end{equation}
for some vector potential $\vect{A}$. We have a gauge freedom here, since $F$ is invariant under the transformation $\vect{A}\rightarrow\vect{A}+\vd\lambda$ for some scalar function $\lambda$, since
\begin{equation}
	\vd\wedge (\vect{A}+\vd\lambda)=\vd\wedge\vect{A}+\vd\wedge(\vd\lambda)
\end{equation}
since $\vd\wedge(\vd\lambda)\equiv 0$
\subsection{Outline of the Paper}
Now I have introduced the main concepts, I begin the paper by decomposing the equation $\vd\vect{z}=0$ grade by grade in order give to give a closed system of equations defining a monogenic function. I call these the, `Clifford-Cauchy-Reimann' equations, or CCR equations for short. \\
I then show that the CCR equations applied to a general multivector in the Spacetime Algebra $Cl(3,1)$ give the Dirac and Maxwell equations. This is not totally surprising-- solutions to the Dirac equation $\vd\phi=0$ are monogenic functions by definition, and it has long been realised that Maxwell's equations can be written as Cauchy- Reimann equations in quaternion analysis (called regularity conditions) \cite{Typaldos1975} and also via complexified quaternions (or biquaternions) \cite{Imaeda1976}, and in  complexified Clifford algebras \cite{Ryan1985}. \\
The link between Dirac and Maxwell equations has been studied in other settings too. In the geometric algebra formalism, we can write the Maxwell equations as the Dirac equation $\vd F=\vect{J}$ \cite{Doran2003}. It has been shown by Kravchenko and Shapiro that the time-harmonic Maxwell and Dirac equations are linked as quaternion - valued operators \cite{Kravchenko1995}\cite{Kravchenko2002}, and Picard, Trostorff and Waurick have shown links between the Maxwell equations as the Dirac operator as matrix operators \cite{Picard2017}. These papers  also include sources in the Maxwell and the Dirac systems. \\
The difference between those earlier results and this paper is that this paper works with the real Clifford algebra $Cl(3,1)$-- which has a direct Geometrical meaning as a algebra generated by Minkowski space. Additionally, this paper begins with a general multivector in $Cl(3,1)$ and shows that if we wish it to be monogenic, it must satisfy the (source-free) Maxwell and Dirac equations.  This is a new result which  implies that the Maxwell and Dirac equations are mathematically fundamental in a way linked to the geometry of Minkowski space.  I end the paper with a short discussion of future work.
\section{Clifford-Cauchy-Riemann equations}
I am not aware of an explicit grade-by-grade decomposition of the Cauchy- Reimann equations for a Clifford algebra in the literature, though they must have been derived before in the course of calculations. I present here such a decomposition, which I will use for the main results in this paper.  We write a general $\x$-valued multivector function in $Cl(p,q)$ as 
\begin{equation}
	\vect{z}(\x)=\sum_{i}\f_{i}=f_0(\x)+f_{1i}(\x)e_{i}+f_{2ij}(\x)e_{i}e_{j}+...+f_{n}e_{1}...e_{n}
\end{equation}
where $\f_{i}$ is a blade of grade $i$, $f(\x)$ is a scalar function of $\x$ and $i$ runs from 1 to $n=p+q$. We are interested in monogenic functions, for which
$\boldsymbol{\partial}\vect{z} =0$. This implies that
\begin{equation}
	\sum_{i}\vd\f_{i}=\sum_{i}\Big(\vd\cdot\f_{i+1}+\vd\wedge\f_{i-1}\Big)=0
\end{equation}
equating all terms with the same grades, we find that
\begin{align}
	&\vd\cdot\f_{i+1}=-\vd\wedge\f_{i-1}, \ i\neq\{1, n-1\} \label{eqn: CCR1} \\ 
	&\vd\cdot \f_{1}=0 \label{eqn:CCR2} \\ 
	&\vd\wedge\f_{n-1}=0 \label{eqn:CCR3}
\end{align}
with $\vd \cdot f_{0}\equiv 0$ and $\vd\wedge\f_{n}\equiv 0$ (you can't take the exterior derivative of a top form, or the divergence of a scalar function). These are the generalisation of the Cauchy-Riemann equations to general Clifford Algebras for the Dirac operator $\vd$.  I shall refer to them as the Clifford-Cauchy-Riemann, or CCR, equations.  Note also that the condition $\vd\cdot\f_{i+1}=-\vd\wedge\f_{i-1}$ becomes $\delta f_{i-1}=df_{i+1}$ in the language of differential forms. 
\subsection{Properties of the CCR Equations} \label{sec:propCCR}
The first thing to note about these equations is that the equations for the odd and even parts seperate. Equation (\ref{eqn: CCR1}) links $\f_{n-1}$ and $\f_{n+1}$, whose grades are always both odd, or both even. Then, equation (\ref{eqn:CCR2}) always involves only $\f_{1}$, which is odd;  and (\ref{eqn:CCR3}) only involves $\f_{d-1}$, which is either odd or even depending upon $d$. Therefore the CCR equations split into parts involving only the odd or even components. This simplifies the analysis of these equations, and we shall make use of this below in section \ref{sec:CCRSA}  when analysing the spacetime algebra. \\
We can rewrite the CCR equations using various dualities. If we define $g_{i}=f_{i}I^{-1}$, then they become
\begin{align}\label{eqn:dual}
	&\vd\wedge\g_{n-(i+1)}I^{-1}=-\vd\cdot\g_{n-(i-1)}I^{-1}, \ i\neq\{1, d-1\}  \\ 
	&\vd\wedge \g_{n-1}I^{-1}=0\\ 
	&\vd\cdot\g_{1}I^{-1}=0
\end{align}
Alternatively, we can only take the dual of one side of equation (\ref{eqn: CCR1})  to get
\begin{equation}
	\vd\wedge\f_{i-1}=\vd\wedge\g_{n-(i+1)}I^{-1}
\end{equation}
Note that whilst in general $\f_{i-1}$ and $\g_{n-(i+1)}$ have different grades, when $n$ is even, $\f_{n/2-1}$ and $\g_{n/2-1}$ both have the same grade in the equation, giving
\begin{equation}\label{eqn:dual1}
	\vd\wedge\f_{n/2-1}=\vd\wedge\g_{n/2-1}I^{-1}
\end{equation}
This suggests that mathematically it could be interesting to investigate functions satisfying $\vect{z}=\pm \vect{z}I^{-1}$, which could be another avenue to explore to better understand the properties of solutions to these equations. 
\section{CCR and the Spacetime Algebra} \label{sec:CCRSA}
We now examine the CCR equations for the Spacetime Algebra. Now, a general multivector in the spacetime algebra can be written in the form
\begin{align}\nonumber
	f_{0}+\f_{1}+\f_{2}+\f_{3}+\f_{4}= f_{0}(\x)+f_{1a}(\x)e_{a}+f_{2ab}(\x)e_{a}e_{b}\\
	+f_{3abc}(\x)e_{i}e_{j}e_{k}+f_{4}(\x)e_{0}e_{1}e_{2}e_{3}
\end{align}
This leads to the CCR equations
\begin{align} \nonumber
	&\vd\cdot\f_{1}=0\\ \nonumber
	&\vd f_{0}=-\vd\cdot \f_{2}\\ \nonumber
	&\vd\wedge\f_{1}=-\vd\cdot\f_{3}\\ \nonumber 
	&\vd\wedge\f_{3}=0 \\ 
	&\vd\cdot\f_{4}=-\vd\wedge\f_{2}
\end{align}
As discussed in section \ref{sec:propCCR}, these split into odd and even graded sectors. 
\subsection{The Odd Sector}
We start with the odd sector
\begin{align}
	&\vd\cdot\f_{1}=0 \label{eqn:odd0}  \\ &\vd\wedge\f_{1}=-\vd\cdot\f_{3} \label{eqn:odd2}  \\& \vd\wedge\f_{3}=0 \label{eqn:odd4}
\end{align}
Since n=4 is even, we can use duality as in equation (\ref{eqn:dual1}) to write the second of these equations as
\begin{equation}\label{eqn:oddsep2}
	\vd\wedge\f_{1}=-\vd\wedge\g_{1}I^{-1}
\end{equation}
where $g_{3}=f_{1}I^{-1}$. To solve this equation, we note that in we first of all note that we can write the electromagnetic field strength in Clifford Algebra form as $\vd\wedge\A=F_{ab}=E_{0i}+(IB)_{jk}$, where $E,B$ are the electric and magnetic fields respectively \cite{Doran2003}. The electric field $E_{0i}$ is a timelike bivector, and the magnetic field $(IB)_{jk}$ is a spacelike one. 
This shows that any vector function $\A$ such that $\vd\wedge\A$ is an anti-selfdual field strength $\vect{F}=-\vect{F}I^{-1}$ will satisfy  equation (\ref{eqn:oddsep2}). Hence $f_{1}=g_{1}=\A$ is always a solution to equation (\ref{eqn:odd2}) since $\vd\wedge\A=-\big(\vd\wedge\A\big)I^{-1}$. \\
We now have a single solution. What are the other possibilities? Suppose that $\vd\wedge\A=-\big(\vd\wedge\vect{C}\big)I^{-1}$ for $\vect{C}\neq\A$.Then $-\big(\vd\wedge\vect{C}\big)I^{-1}-\big(\vd\wedge\vect{A}\big)I^{-1}=0$, which implies that $\vd\wedge(\vect{C}-\A)=0$, and therefore $\vect{C}$ and $\A$ differ by the gradient of a scalar function, since $\vd\wedge\vd\lambda=0$ automatically.  Then the most general solution we can write is 
\begin{align}
	\f_{1}&=\A+\vd \lambda_{1},\\
	\f_{3}&=g_{1}I^{-1}=\big(\A+\vd\lambda_{2}\big)I^{-1}
\end{align}
for scalar functions $\lambda_{1}$ and $\lambda_{2}$. If we set $\lambda_{1}= \lambda_{2}\equiv\lambda$, and write $\tilde{\A}$=$\A+\vd\lambda$ then equation (\ref{eqn:oddsep2}) becomes the equation for a single anti-selfdual Field Strength. 
\begin{equation}\label{eqn:odd2sol}
	\vd\wedge\tilde{\A}=\big(\vd\wedge\tilde{\A}\big)I^{-1}
\end{equation}
 To confirm that $\vect{F}$ really is an electromagnetic field strength tensor, note that if we write $\vect{F}=\vd\wedge\tilde{\A}$ then
\begin{align} \nonumber
	\vd\vect{F}&=\vd\cdot\vd\wedge\tilde{\A}+\vd\wedge\vd\wedge\tilde{\A}\\ \nonumber
	&=\vd\cdot\vd\wedge\tilde{\A}=-\vd\cdot\big(\vd\cdot(\A I^{-1})\big)=0
\end{align} 
which is simply the equation for a source-free Maxwell field written in Geometric Algebra notation \cite{Doran2003}. I have used the fact that $\vd\cdot\vd\cdot \f=\vd\wedge\vd\wedge \f=0$ for any $\f$ in the second and fourth inequalities, and equation (\ref{eqn:odd2sol}) in the second. \\
We can now look at the equation $\vd\cdot\f_{1}=0$. This is Gauss' law
\begin{equation}
	\vd\cdot\tilde{\A}=0
\end{equation}
Note that if we extract the gauge function $\lambda$ we get
\begin{equation}
	\vd\cdot\A+\vd^{2}\lambda=0
\end{equation}
which allows us to use $\lambda$ to set a gauge exactly as in standard presentations of electromagnetism\cite{Manton2004}. \\
What about the remaining equation $\vd\wedge\f_{3}=0$? By the solution in equation (\ref{eqn:odd2sol}) this  is equivalent to $\vd\wedge(\tilde{A}I^{-1})=0$ which implies that $(\vd\cdot\tilde{\A})I^{-1}=0$, which implies that $\vd\cdot\A+\vd^{2}\lambda=0$, just the same as equation (\ref{eqn:odd0}).\\ 
Putting it all together, we see that the odd sector of the CCR equations is
\begin{align} 
	\vd\wedge\tilde{\A} & =\big(\vd\wedge\tilde{\A}\big)I^{-1} \label{eqn. asd} \\
	\vd\cdot\tilde{\A}&=0 \label{eqn. Gauss}
\end{align}
This describes an anti-selfdual field arising from a vector potential $\tilde{A}$ (equation (\ref{eqn. asd}), which obeys Gauss' Law (equation (\ref{eqn. Gauss}). We also have a gauge freedom to rescale $\tilde{A}$ by a scalar function $\lambda$, via $\tilde{A}\rightarrow\tilde{A}+\vd\lambda$.\\
One mystery remains. In the above derivation, I assumed that $\lambda_{1}=\lambda_{2}=\lambda$. What about when $\lambda_{1}\neq\lambda_{2}$? In this case we have
\begin{equation}
	f_{1}=\A+\vd\lambda_{1}, \ \ \ \ \ f_{3}=\big(\A+\vd\lambda_{2}\big)I^{-1}
\end{equation}
Putting this into equations (\ref{eqn:odd0}) and (\ref{eqn:odd4}) we get that
\begin{equation}\label{eqn:lambda1}
	\vd\cdot\A=-\vd^{2}\lambda_{1}, \ \ \ \vd\cdot\A=-\vd^{2}\lambda_{2}
\end{equation}
This implies that $\vd^{2}\big(\lambda_{1}-\lambda_{2}\big)=0$, and therefore that $\lambda_{1}-\lambda_{2}$ is a harmonic function. I am unsure of the physical significance of this, however I hope to explore this in future work. 
\subsection{The Even Sector}
The CCR equations for the even subalgebra are given by
\begin{align}
	\vd f_{0}=-\vd\cdot\f_{2} \label{eqn:even1} \\
	\vd \wedge \f_{2}=-\vd\cdot\f_{4} \label{eqn:even3}
\end{align}
Given that the even subalgebra is $Spin(3,1)$, we should expect that these equations are linked to that of a massless spinor.  Written in the Geometric Algebra representation, this is 
\begin{equation}\label{eqn:FreeDirac}
	\vd\phi=0
\end{equation}
where $\phi=\rho^{1/2} e^{IB} e^{\boldsymbol{\theta/2}}$. Here, $\phi(x)$ is a multivector function, $\rho(\x)$ and $B(\x)$ are scalar functions, and $\tht(\x)$ is a bivector function. This form of the solution to Dirac's equation is due to David Hestenes \cite{Hestenes2003}.  Mathematically this corresponds to a polar decomposition of $\phi$. There are eight degrees of freedom, just as we would expect -- one each for $\rho$ and $B$, and six contained in $\theta$, which generates a Lorentz rotation. Physically, following Hestenes, we can interpret the multivector function $\phi$ as a physical wave in Minkowski space, with $\rho^{1/2}$ as the amplitude, and $e^{\tht/2}$ being the spinor generator of a rotation into the rest frame of the particle. The physical interpretation of $B$ is more ambiguous, but Hestenes has suggested that for the full Dirac equation it corresponds to a hypothetical rapid oscillation of the electron called Zitterbewegung \cite{Hestenes2010}. I will not address these interpretational issues here. \\
Returning to equation (\ref{eqn:FreeDirac}), we can evaluate
\begin{equation}
	\vd\phi=\Big(\frac{\partial\rho}{\rho}+\vd(IB)+\vd\tht\Big)\rho^{1/2}e^{IB}e^{\tht/2}=0
\end{equation}
Where as above, $\phi(x)$ is a multivector function, $\rho(\x)$ and $B(\x)$ are scalar functions, and $\tht(\x)$ is a bivector function. This implies 
\begin{equation}
	\frac{\partial\rho}{\rho}+\vd(IB)+\vd\theta/2=0
\end{equation}
collecting terms of the same grade, we find
\begin{equation}
	\vd\text{ln}(\rho)=-\vd\cdot\tht/2
\end{equation}
for grade 1, and 
\begin{equation}
	\vd\wedge\tht/2=-\vd\cdot(IB)
\end{equation}
for grade 3, with $\vd\wedge(IB)\equiv0$ identically. But these are just the CCR equations (\ref{eqn:even1}) and (\ref{eqn:even3}), with $f_{0}=\text{ln}\rho$, $\f_{2}=\tht/2$ and $\f_{4}=IB$. Therefore, $f_{0}, \f_{2}, \f_{4}$ satisfying the CCR equations automatically define a free Dirac field via (\ref{eqn:FreeDirac}). \\
A final note: We could have written the scalar part of $\phi$ as $e^{a(\x)/2}$ for some scalar function $a(\x)$, rather than using $\rho^{1/2}$. This would have given us $f_{0}=a/2$, rather than $\text{ln}(\rho)$. I chose the notation $\rho^{1/2}$ partly to fit with the notation of Hestenes \cite{Hestenes2003}, and partly because of the similarity of the term $\frac{\vd\rho}{\rho}$ to the quantum potential of Bohm \cite{Bohm1980}, which is derived from a similar polar decomposition of the wavefunction. 
\section{Conclusion and Further Work}
Putting it all together, we find that a multivector
\begin{equation}
	\vect{z}=\text{ln}(\rho)+\tilde{\A}+\tht/2+\tilde{\A}I^{-1}+IB
\end{equation}
satisfying the CCR equations in the Spacetime Algebra $Cl(3,1)$ defines both a free Dirac field $\phi=\rho^{1/2}e^{IB}e^{\tht/2}$, and an electromagnetic field strength $\vect{F}=\vd\wedge\tilde{\A}$, where $\tilde{\A}$ satisfies Gauss' law, and is defined up to the addition of a gauge  $\vd\lambda$. This shows that the Dirac and Maxwell equations are not separate or arbitrary, but are the four dimensional analogue of the Cauchy- Riemann equations in Complex Analysis, and hence deeply linked to the Clifford Geometry of Spacetime. It also implies that Clifford Analysis will be a powerful tool for theoretical physics as it is developed further. \\
The obvious next question is how to derive the equations for a massive or charged Dirac field, or for an electromagnetic field with a source $\vect{J}=\vd\vect{F}$? I suspect that this would involve considering  Dirac operators of the form $\vd+\vect{A}$, where $\vect{A}$ is a gauge potential.  It would be useful here to engage with the works mentioned in the introduction, by Kravchenko et.al, and Picard et.al. \cite{Kravchenko2002}\cite{Kravchenko1995}\cite{Picard2017}. \\
It is also worth noting here that Hiley and Callaghan \cite{Hiley2010} \cite{Hiley2011} have shown that a general multivector in Clifford Algebra can be used to define a quantum wavefunction. This is another angle to explore. 
A second direction for research is given by the observation that, since solutions to the Dirac equation are in one-to-one correspondence with Harmonic functions, the spaces of monogenic functions are (at least for compact manifolds) determine the cohomology of the underlying manifold. (This is due to Hodge Theory \cite{Schwarz1995}, which states that the spaces of  harmonic forms on a manifold are isomorphic to the de Rham cohomology groups ). This implies a strong link between topology and solutions to the Maxwell and Dirac equations. \\
Of course we are interested in Lorentzian manifolds; and the cohomology of Lorentzian spaces is notoriously difficult  \cite{khavkine2016}\cite{Benini2016}\cite{Baer2015} . I suspect, however, that the approach in this paper may yield new results.  One strategy , once we have extended the theory to general Dirac operators, might be to look at the  BRST or BV cohomologies  \cite{Henneaux1994}\cite{Rejzner2020}, making use of the fact that Geometric Algebra allows us to give a direct geometric interpretation of all our expressions to determine the physical meaning of the resulting cohomology groups. \\

\bibliographystyle{spmpsci}
\bibliography{Jan2023}

\end{document}